\documentclass[a4paper]{jpconf}
\usepackage{graphicx}
\graphicspath{{./},{/home/user/fig/publi/muon/}}

\begin{document}

\title{Low temperature crystal structure and local magnetometry for the geometrically frustrated 
pyrochlore Tb$_2$Ti$_2$O$_7$}

\author{P.~Dalmas~de~R\'eotier$^{1,2}$, A.~Yaouanc$^{1,2}$, A.~Bertin$^{1,2}$, C.~Marin$^{1,2}$, S.~Vanishri$^{1,2}$, D. Sheptyakov$^3$, A.~Cervellino$^4$, B.~Roessli$^3$, C.~Baines$^5$}

\address{$^1$Univ. Grenoble Alpes, INAC-SPSMS, F-38000 Grenoble, France}
\address{$^2$CEA, INAC-SPSMS, F-38000 Grenoble, France}
\address{$^3$Laboratory for Neutron Scattering and Imaging, Paul Scherrer Institute, CH-5232 Villigen-PSI, Switzerland}
\address{$^4$Swiss Light Source, Paul Scherrer Institute, CH-5232 Villigen-PSI, Switzerland}
\address{$^5$Laboratory for Muon-Spin Spectroscopy, Paul Scherrer Institute, CH-5232 Villigen-PSI, Switzerland}


\begin{abstract}

We report synchrotron radiation diffraction and  muon spin rotation ($\mu$SR) measurements on the frustrated pyrochlore magnet
Tb$_2$Ti$_2$O$_7$. The powder diffraction study of a crushed crystal fragment does not reveal any 
structural change down to $4$~K. The $\mu$SR measurements performed at 20~mK on a mosaic of single 
crystals with an external magnetic field applied along a three-fold axis are consistent with 
published a.c.\ magnetic-susceptibility measurements at 16~mK. While an inflection point could be 
present around an internal field intensity slightly above 0.3~T, the data barely support the presence 
of a magnetization plateau.
\end{abstract}



The study of geometrically frustrated magnetic systems reveals a large variety of new magnetic phases.
Among the frustrated materials, the rare-earth pyrochlore oxides which crystallize in a cubic 
structure ($Fd\bar{3}m$ space group) form a family for which different exotic ground states have been 
found \cite{Gardner10}. Focusing on the insulators, we mention (i) the spin-ice ground state of 
Ho$_2$Ti$_2$O$_7$ and Dy$_2$Ti$_2$O$_7$ \cite{Harris97,Ramirez99}, (ii) the  spin-liquid ground state 
of Yb$_2$Ti$_2$O$_7$ \cite{Hodges02} characterized by a pronounced peak in the specific heat for 
powder samples \cite{Yaouanc11c,Ross11a}, although an exotic ordered magnetic state has been reported for a 
single crystal \cite{Yasui02,Chang12}, (iii) the unconventional dynamical ground state of
Tb$_2$Sn$_2$O$_7$ for which magnetic Bragg reflections are observed by neutron 
diffraction \cite{Mirebeau05} while no spontaneous magnetic field is found by the zero-field 
positive muon spin relaxation ($\mu$SR) 
technique \cite{Dalmas06,Bert06}, (iv) the persistent spin dynamics detected in the ordered states of 
Gd$_2$Sn$_2$O$_7$, Gd$_2$Ti$_2$O$_7$ and 
Er$_2$Ti$_2$O$_7$ \cite{Bertin02,Yaouanc05a,Chapuis09b,Lago05,Dalmas12a}, and (v)
the splayed ferromagnet Yb$_2$Sn$_2$O$_7$, i.e.\ essentially a ferromagnetic 
compound \cite{Yaouanc13,Dun13,Lago14}, with an emergent gauge field \cite{Savary12}.

The most mysterious compound of the rare-earth pyrochlore oxide family might be Tb$_2$Ti$_2$O$_7$ for which 
no long-range magnetic order is detected down to 20~mK, far below the absolute value of its Curie-Weiss 
temperature $\Theta_{\rm CW} = - 19$~K \cite{Gardner99,Yaouanc11a}. Two theoretical ground states
have been suggested: a quantum spin ice ground state proposed in Ref.~\cite{Molavian07} and a 
Jahn-Teller like distorted ground state based on specific heat measurements \cite{Chapuis10}.
In analogy with the spin-ice systems, a magnetization plateau is expected at low temperature for 
an external magnetic field ${\bf B}_{\rm ext}$ applied along a $[111]$ crystal direction if the 
former ground state is reached \cite{Molavian09}. While static magnetization measurements have not 
found any signature of the predicted plateau down to 43~mK \cite{Lhotel12,Legl12,Sazonov13},
the presence of a weak magnetization plateau below about 0.05~K has been proposed from
a.c.\ susceptibility experiments for 
$0.06 < B_{\rm ext} <0.6$~T \cite{Yin13}. Pointing out to the complexity of the physics involved, 
an anomaly in the static magnetic response has been observed around
0.15~K \cite{Luo01,Yaouanc11a,Lhotel12,Legl12} which strangely enough corresponds to a minimum 
rather than a maximum in the 
specific heat \cite{Yaouanc11a}. Pinch points in the neutron scattering intensity have been
observed \cite{Fennell12} with dispersive excitations emerging from them \cite{Guitteny13,Fennell14},
suggesting a strong magnetoelastic coupling in the Coulomb phase of Tb$_2$Ti$_2$O$_7$.
In addition, a neutron scattering intensity measured at the $\left ( {1 \over 2}, {1 \over 2}, {1 \over 2} \right )$ 
position in 
reciprocal space has attracted some attention \cite{Fennell12,Petit12,Fritsch13,Taniguchi13}.
However, its characterization and origin are a subject of discussions.

Given the dependence of some physical properties of Tb$_2$Ti$_2$O$_7$ 
on the sample preparation \cite{Chapuis10,Chapuis09a,Yaouanc11a,Ross11a,Lhotel12,Taniguchi13},
here we shall first present a high-resolution synchrotron radiation diffraction study on a crushed 
fragment of a single crystal. Thereafter we shall report on frequency shifts as measured on a crystal by the 
transverse-field (TF)-$\mu$SR technique with an external field ${\bf B}_{\rm ext}$ applied 
along a three-fold axis. They give access to the thermal and field dependences of a physical quantity 
proportional to the static magnetization. These measurements, for which the influence of a small amount of an impurity phase, if any, should be minimum,\footnote{As an example, see A. Yaouanc {\it et al}, to be published.}
barely support the existence of magnetization plateau at 20~mK.

For the two experiments we used parts of the crystal denoted C in 
Refs.~\cite{Chapuis10,Chapuis09a,Lhotel12}. This crystal is characterized by a very small residual 
entropy \cite{Yaouanc11a}, as expected for a sample in thermodynamical equilibrium \cite{Revell13}. We briefly 
recall the procedure for the crystal synthesis. A polycrystalline Tb$_2$Ti$_2$O$_7$ powder has been first 
prepared from commercial Tb$_2$O$_3$ and TiO$_2$ powders of respective purity 4N and 4N5. 
An initial heat treatment of these powders to 1200$^\circ$C has been 
followed by a second treatment up to 1350$^\circ$C with an intermediate grinding and compaction so 
as to obtain a dense rod. The crystal has been subsequently prepared from the  Tb$_2$Ti$_2$O$_7$ 
rod using the traveling floating zone technique under oxygen gas and with a translational velocity 
of 7 mm/h. For the $\mu$SR measurements, plates whose normal axis is a [111] axis have been cut from the 
single crystal: their thickness is about 1/3~mm and their lateral size is up to 6~mm. 

The synchrotron radiation measurements have been performed at the high resolution powder diffractometer
of the Material Science (MS) beamline at the Swiss Light Source (Paul Scherrer Institut, Switzerland).
An x-ray beam of wavelength 0.49646\,\AA, corresponding to an energy $\simeq 25.0$~keV, has been used. The 
x-ray flux is maximum at this energy at the MS beamline \cite{Willmott13}. A crushed fragment
of the Tb$_2$Ti$_2$O$_7$ crystal and $\simeq 15 \, {\rm wt.} \%$ of silicon powder have been mixed and 
ground to obtain a homogeneous mixture. The specimen has been loaded into a $0.3$~mm diameter glass capillary. 
The presence of silicon helps in reducing the Tb$_2$Ti$_2$O$_7$ sample x-ray absorption and provides a convenient calibration. The data have been taken from room temperature down to $4$~K. 

\begin{figure}
\begin{minipage}[t]{3.5in}
\centering
\includegraphics[height=2.1in]{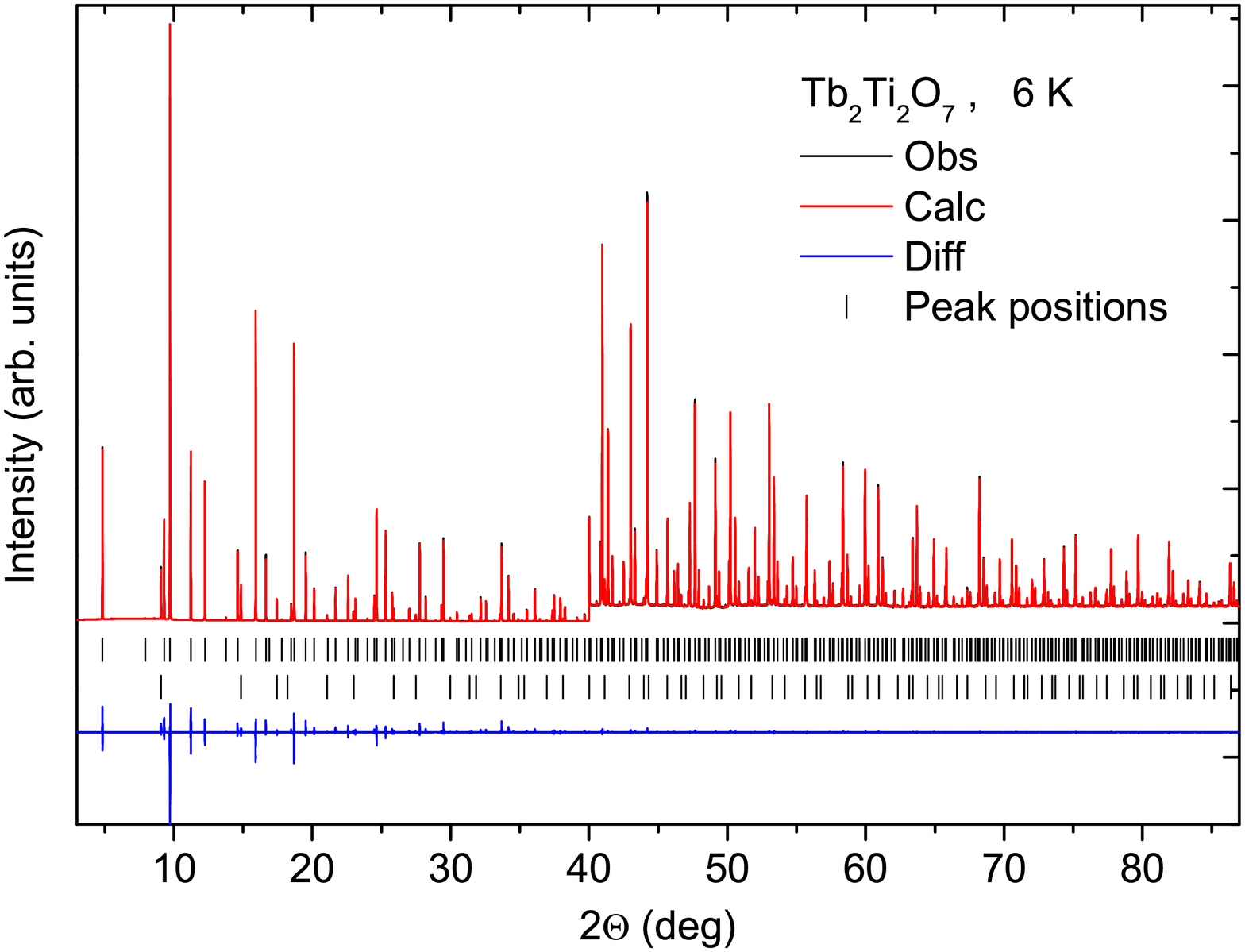}
\caption{
Rietveld refinement plot of synchrotron x-ray powder diffraction data of Tb$_2$Ti$_2$O$_7$ 
collected at 6~K with a photon energy of 25~keV. Observed points, calculated profile 
and difference curve are shown. Ticks below the graph show the calculated peak 
positions for Tb$_2$Ti$_2$O$_7$ and Si (upper and lower rows respectively). The intensities 
beyond $2\Theta$ = 40$^\circ$ have been enlarged by a factor of 10 in order to illustrate 
the agreement quality at higher angles. 
}
\label{diagram_diffraction_6K}
\end{minipage}
\hfill
\begin{minipage}[t]{2.5in}
\centering
\includegraphics[height=2in]{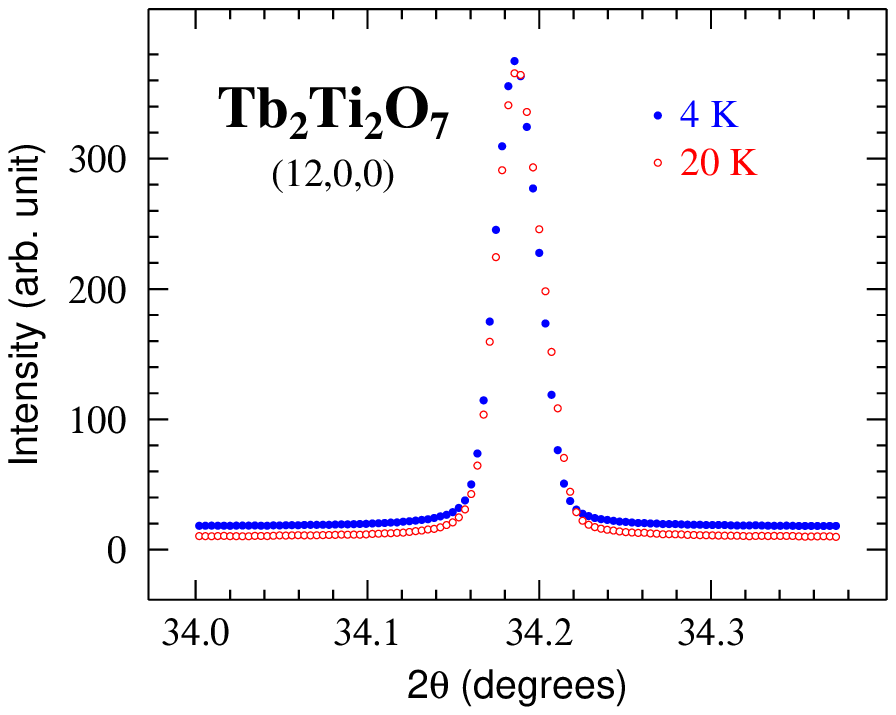}
\caption{
Comparison of the $(12,0,0)$ x-ray Bragg peak profiles measured at 20 and 4~K for our
Tb$_2$Ti$_2$O$_7$ powder. The full width at half maximum of the Bragg peak is $9\times 10^{-3}$ in 
reciprocal lattice units.
}
\label{comparison1200}
\end{minipage}
\end{figure}

\begin{figure}[tb]
\begin{minipage}[t]{3in}
\centering
\includegraphics[height=2in]{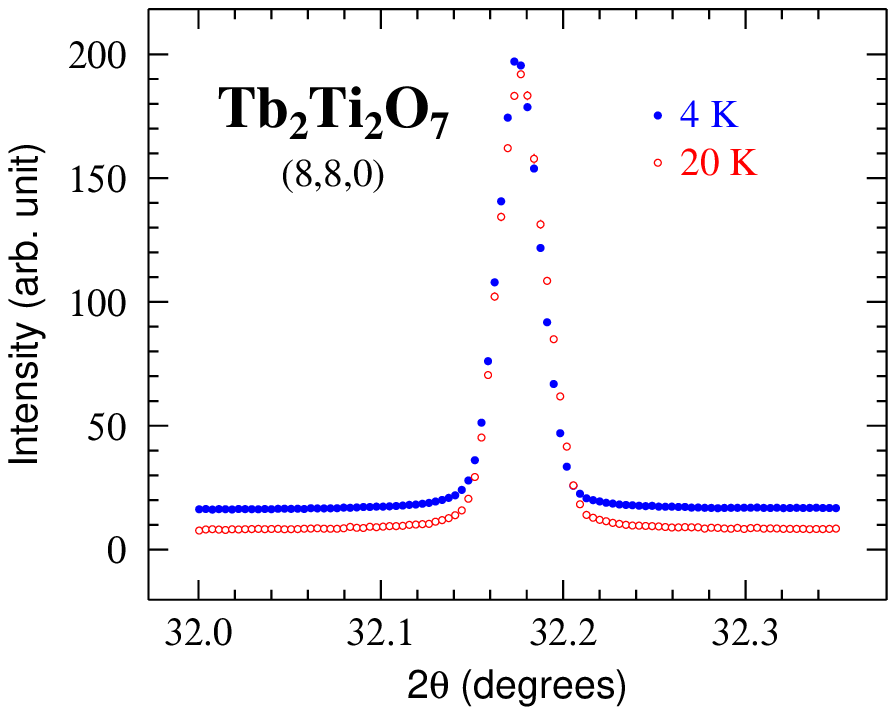}
\caption{
Comparison of the $(8,8,0)$ x-ray Bragg peak profiles measured at 20 and 4~K for our
Tb$_2$Ti$_2$O$_7$ powder. 
}
\label{comparison880}
\end{minipage}
\hfill
\begin{minipage}[t]{3in}
\centering
\includegraphics[height=2in]{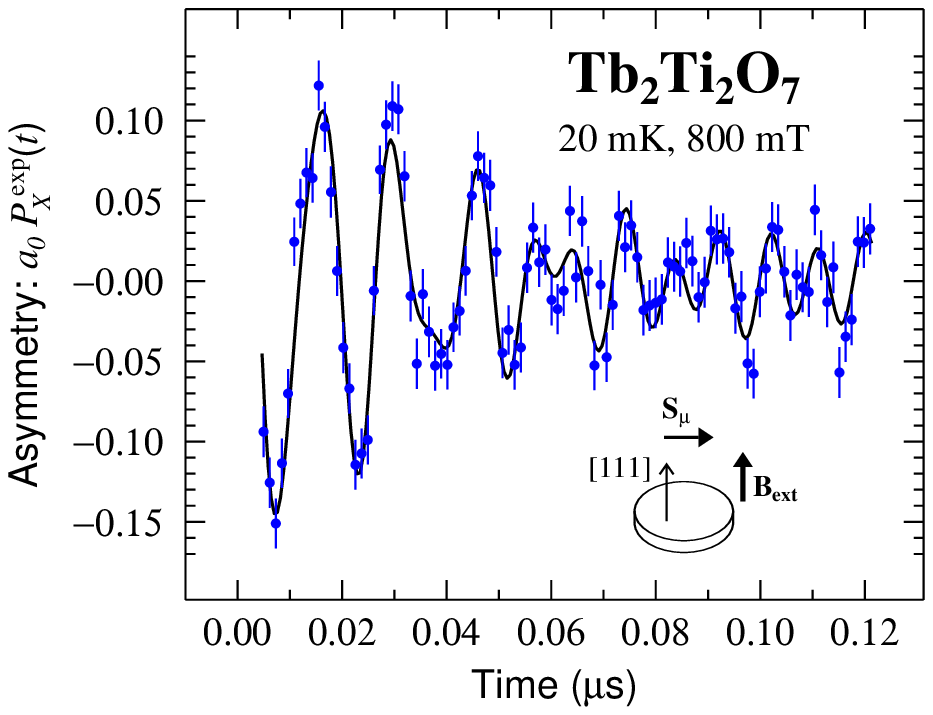}
\caption{A typical TF-$\mu$SR asymmetry time spectrum recorded at 20~mK for a
mosaic of Tb$_2 $Ti$_2$O$_7$ crystals with ${\bf B}_{\rm ext}$ 
applied along a three-fold axis and $B_{\rm ext} = 800$~mT. The solid line results from a fit as explained
in the main text.}
\label{muon_spectra}
\end{minipage}
\end{figure}

Figure~\ref{diagram_diffraction_6K} 
displays an example of a diffraction diagram. The solid line results from a Rietveld fit. The 
quality of the fit is excellent. In Figs.~\ref{comparison1200} and \ref{comparison880} we compare the profiles 
of the $(12,0,0)$  and $(8,8,0)$ Bragg profiles recorded at $20$ and $4$~K. 
In contrast to the result of Ruff {\it et al.} \cite{Ruff07} who found a profile broadening
below $20$~K, no shape change of the profiles is observed at low temperature. To further 
check for the behavior of our Tb$_2$Ti$_2$O$_7$ specimen, 
we determined the temperature dependence of the lattice parameter $a$. The compound exhibits the expected 
smooth thermal contraction as it is cooled down with a plateau below $\approx 25$~K to $a\simeq$ 10.1368~\AA. 
In Ref.~\cite{Ruff07} an expansion was reported below $20$~K. Therefore, our measurements do not support 
the previous claim of the presence of structural fluctuations below $\approx 20$~K in Tb$_2$Ti$_2$O$_7$. They 
are in fact fully consistent with the results of Goto {\it et al.} \cite{Goto12}. 
We measure for the lattice parameter $a= 10.15735 \, (10)$~{\AA} at 295~K. This is slightly
larger than  $a= 10.15529 \, (1)$~{\AA} recently reported \cite{Fennell14}. In fact, as pointed 
out in this reference, literature values of the lattice parameter of Tb$_2$Ti$_2$O$_7$
are clustered around 10.154~\AA, but outlying values do exist. This casts doubt on the 
assignment of composition based on the lattice parameter \cite{Taniguchi13}.

Finally the value for the position parameter of the oxygen ion at position  $48f$ is 
$x$ = 0.32783\,(3) for temperatures below 50~K.\footnote{The $x$ values are given for the description of the space group with the 
origin at site of symmetry $.\overline{3}m$. The origin of the lattice is occupied by a 
titanium atom.} Above 200~K we find that it slightly decreases: 
$x$ = 0.3270\,(1). These values are in agreement with the literature \cite{Helean04}.

The $\mu$SR study has been performed at the Low Temperature Facility of the Swiss Muon Source 
(Paul Scherrer Institute, Switzerland) with a disk made of a mosaic of crystal plates previously
described. It has covered the temperature range from 20 to 500~mK. We have set ${\bf B}_{\rm ext}$ 
perpendicular to the sample disk and therefore parallel to a Tb$_2$Ti$_2$O$_7$ $[111]$ axis, and 
the initial muon polarization ${\bf S}_\mu$ perpendicular to it. By definition ${\bf B}_{\rm ext}$ 
and ${\bf S}_\mu$ are parallel to the $Z$ and $X$ axes, respectively.
The geometry of the experiment is illustrated in the inset of Fig.~\ref{muon_spectra}.

The quantity of interest is the TF-$\mu$SR asymmetry time spectrum $a_0 P^{\rm exp}_X(t)$, 
where $P^{\rm exp}_X(t)$ describes the evolution of the muon polarization under 
${\bf B}_{\rm ext}$.  The positron detectors are located perpendicular to the $Z$ axis \cite{Yaouanc11}. 
In Fig.~\ref{muon_spectra}
a typical asymmetry spectrum is displayed. It can be described as the weighted sum of two 
beating damped oscillating components: the first accounting for the muons implanted in the 
sample and the second modeling the response for the muons stopped in the sample 
surroundings, essentially the silver sample holder:
\begin{eqnarray}
a_0 P^{\rm exp}_X(t) & = & a_1\exp(-\lambda_{X,1} t)\cos(2\pi \nu_1 t + \varphi) +\, a_2  \exp(-\lambda_{X,2} t)\cos(2\pi \nu_2 t + \varphi).
\label{Fit_1}
\end{eqnarray}
The analysis of the  presented spectrum yields $a_1 = 0.192 \, (13)$ and
$a_2 = 0.028 \, (2)$, i.e.\ only $ \approx 13 \%$ of the muons are stopped
in the surroundings with $\nu_2 = 108.46 \, (1)$~MHz. This latter value is very close to the
precession frequency $\nu_{\rm ext} = \gamma_\mu B_{\rm ext}/(2 \pi)=108.43$~MHz expected for muons 
submitted to a field $B_{\rm ext} = 800$~mT. Here $\gamma_\mu$ = 851.6~Mrad\,s$^{-1}$\,T$^{-1}$ is the 
muon gyromagnetic ratio.

\begin{figure}[tb]
\begin{minipage}[t]{3in}
\centering
\includegraphics[height=2in]{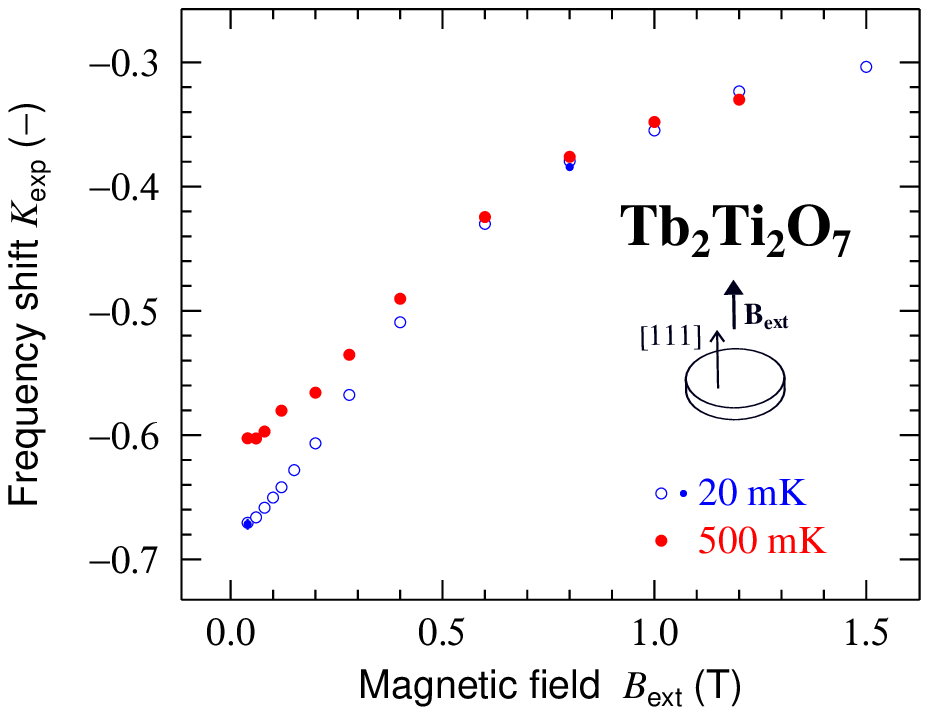}
\caption{Field intensity dependence of the normalized frequency shift $K_{\rm exp}$ for our mosaic of 
Tb$_2$Ti$_2$O$_7$ crystals. ${\bf B}_{\rm ext}$ has been applied along a three-fold axis. The shifts 
have been measured at 20 and 500~mK. The error bars are smaller than the symbols. The data at 20~mK 
shown by open circles have been measured after zero-field cooling by increasing $B_{\rm ext}$ up to
1.5~T. Further data (small bullets) recorded after decreasing $B_{\rm ext}$ to 800 and 40~mT show 
no hysteretic effect. In contrast, an hysteretic response at $60$~mT and low temperature has been 
observed in a temperature scan \cite{Yaouanc11a}.}
\label{shift_B}
\end{minipage}
\hfill
\begin{minipage}[t]{3in}
\centering
\includegraphics[height=2in]{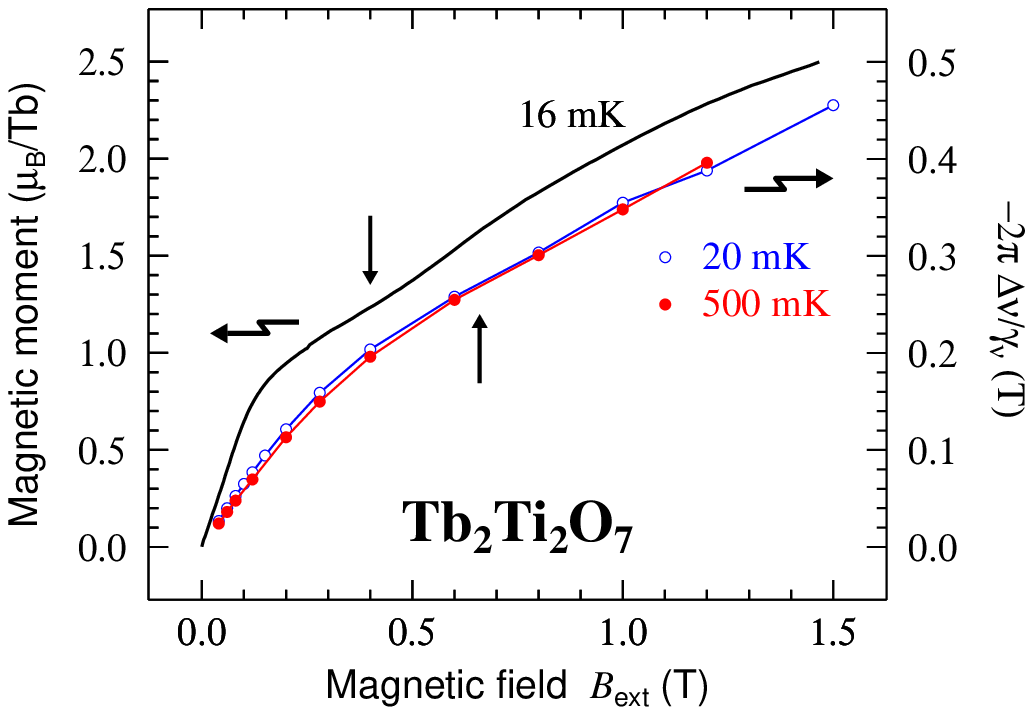}
\caption{Open circles and bullets: product 
$- B_{\rm ext} K_{\rm exp} = -2 \pi \Delta \nu/\gamma_\mu$ deduced from the 20 and 500~mK data displayed 
in Fig.~\ref{shift_B}, versus $B_{\rm ext}$. The experimental points are linked by segments. Solid 
line: field dependence of the terbium magnetic moment measured at 16~mK for ${\bf B}_{\rm ext}$ applied 
along a [111] crystal direction. As explained in the main text, the latter curve is computed from the 
data published by Yin {\it et al\/} \cite{Yin13}.}
\label{B_times_shift_B}
\end{minipage}
\end{figure}

The interest is not in $\nu_1$, but rather in the frequency shift
$\Delta \nu = \nu_1 - \nu_2$, or more conveniently the normalized frequency shift 
$K_{\rm exp} = \Delta \nu/ \nu_2$ which, to a good approximation, can be written as
$K_{\rm exp} = \Delta \nu/ \nu_{\rm ext}$. This quantity, already known to be 
negative for Tb$_2$Ti$_2$O$_7$ \cite{Ofer07,Yaouanc11a}, is displayed in Fig.~\ref{shift_B} 
versus $B_{\rm ext}$ for $T = 500$ and $20$~mK. Relative to the data at 500~mK, at 20~mK an extra 
contribution to $K_{\rm exp}$ appears below $ \approx 0.6$~T. This is consistent with the
first magnetization curves recorded at 500 and 57~mK which also differentiate
themselves only below $0.5$~T \cite{Lhotel12}. Following the difficulties recently pointed out
to correct for demagnetization \cite{Bovo13}, we shall refrain to do it for our data.
However, the qualitative consistency of our data and the magnetizations displayed in 
Ref.~\cite{Lhotel12}, for which demagnetization corrections were not required, suggests
our field scale to be correct.

Since Tb$_2$Ti$_2$O$_7$ is an insulator, only the dipole field from the terbium magnetic moments
contributes to the frequency shift. For the sake of derivation, assuming the sample to be a sphere, i.e.\
the relevant demagnetization coefficient $N^Z = 1/3$, 
$K_{\rm exp} = K^\prime_{\rm dip} = {\mathcal D}^{ZZ} (\mu_0 M/B_{\rm ext})$ where 
$K^\prime_{\rm dip}$ accounts for the shift from the dipole field arising from the magnetic moments 
inside the Lorentz sphere, ${\mathcal D}^{ZZ}$ is a dipole tensor element, and $M$ is the 
magnetization along a three-fold axis. Note that if we were in the linear regime of the magnetization, the term in parentheses would be the 
magnetic susceptibility $\chi$ and we 
would recover the usual formula \cite{Yaouanc11} $K_{\rm exp} = K^\prime_{\rm dip} = {\mathcal D}^{ZZ}\chi$.\footnote{Here we have neglected the anisotropy of the susceptibility.} Hence, we expect the product $- B_{\rm ext} K_{\rm exp}$ 
to be proportional to $M$. In Fig.~\ref{B_times_shift_B} we plot this product versus $B_{\rm ext}$.
If there was a definitive plateau in the magnetization, the product would be field independent in 
a finite field range. This is not observed. However, as indicated by the up-arrow, a weak inflection
point is present for the 20~mK data at $B_{\rm ext} \simeq 0.66$~T. It has disappeared at 500~mK.

Yin {\it et al.} have performed a.c.\ magnetic-susceptibility measurements for a Tb$_2$Ti$_2$O$_7$ 
crystal \cite{Yin13} with ${\bf B}_{\bf ext}$ parallel to a three-fold axis. 
The data presented in their Fig.~1d for 
$T = 16$~mT have been recorded with an a.c.\ field amplitude of 0.94~mT and $B_{\rm ext}$ extending 
up to 1.5~T. With these experimental conditions the measurements give access to 
$\mu_0{\rm d} \chi^\prime_Z /{\rm d} B_{\rm ext}$ rather than $\chi^\prime_Z$, where $\chi^\prime_Z$ is 
the real part of the susceptibility. This means that $M$ can be deduced from the data of 
Yin {\it et al.} by field integration. Normalizing the result by the volume per terbium ion 
and dividing by the electronic Bohr magneton, we obtain the magnetic moment per terbium ion versus 
$B_{\rm ext}$ displayed in Fig.~\ref{B_times_shift_B}. An inflection point occurs at 
$B_{\rm ext} \simeq 0.40$~T as indicated by the down-arrow. In Fig.~1d of Ref.~\cite{Yin13} the data were not 
corrected for the demagnetization field. However, it is relatively small: about $60$~mT around $0.6$~T; see 
Note 35 in Ref.~\cite{Yin13}. Because of our experimental geometry, see Fig.~\ref{shift_B}, it is 
expected to be larger in our case. Assuming a reasonable demagnetization field of $\approx 0.30$~T 
at $0.66$~T, the position of the inflection point inferred from our data and from Ref.~\cite{Yin13} 
would coincide. Expressed in terms of the internal magnetic field, the inflection
point would be located slightly above 0.3~T.

Therefore our $\mu$SR measurements and a.c.\ susceptibility data suggest an inflection point in the
magnetization could be present slightly above $0.3$~T at low temperature. As deduced from 
Fig.~\ref{B_times_shift_B}, there is an apparent discrepancy between the two data sets at lower field.
However, since the initial slope of the magnetization curve at $57$~mT is about 
$10 \, \mu_{\rm B}/{\rm T}$ \cite{Lhotel12}, we compute a susceptibility of $\approx 1.8$, a quite
large value, in agreement with the literature \cite{Yin13}.  Our frequency shift data at low field 
are therefore mainly controlled by the shape of our mosaic sample. From Ref.~\cite{Lhotel12} we 
estimate $\mu_0 M/B_{\rm ext}$ to be about three times smaller at $0.4$~T. This much smaller value
explains that the field dependence of the two data sets are similar at high field.

The authors of the a.c.\ susceptibility measurements have stressed that $\chi^\prime_Z$ displays a bump 
at $ B_{\rm ext} \simeq 0.5$~T for $T = 40$~mK  only if the field is swept sufficiently slowly, i.e\ 
for $0.625$~mT/min but not for $3.75$~mT/min; see Fig.~S2 of the supplemental material to Ref.~\cite{Yin13}. 
Our measurement sequences have consisted of changing $B_{\rm ext}$ step by step, taking between 
$10$ to $20$~min for a field change. This is therefore quicker than 
$0.625$~mT/min. However, when at a given $B_{\rm ext}$ value, we stood there without any field change 
about $1$~hour for data taking. This is in contrast to the a.c.\ susceptibility measurements done with an a.c.\ field of 
frequency 9.6~Hz for the data of Fig.~S2 and 0.21~Hz for the measurements presented in Fig.~1d.

In summary, we have reported a high resolution synchrotron radiation diffraction investigation of a piece
of a crushed Tb$_2$Ti$_2$O$_7$ single crystal, and frequency shift $\mu$SR measurements at 20~mK on a 
mosaic of single crystals of the same compound for an external field applied along a three-fold axis. The
x-ray study has not revealed any structure change down to $4$~K, confirming the high quality of our
crystals. The $\mu$SR measurements have given us access to a quantity proportional to the terbium magnetic 
moment. While an inflection point seems to be present slightly above 0.3~T for the field dependence of the moment at the lowest temperature, in agreement with previously published a.c.\ susceptibility data, there is barely any signature of a 
magnetization plateau.

\section*{Acknowledgments}
PDR gratefully acknowledges partial support of Prof.\ H.\ Keller from the University of Zurich for the
$\mu$SR measurements. Part of this work has been performed at the Swiss Light Source and the Swiss Muon 
Source, Paul Scherrer Institute, Villigen, Switzerland.

\section*{References}
\bibliographystyle{iopart-num-wo-url.bst}
\bibliography{reference}

\end{document}